\documentclass[preprint,showpacs,aps,floatfix]{revtex4}
\usepackage{graphicx}
\usepackage{graphics}
\usepackage{amssymb}
\def\d{{\rm d}}

\begin{document}

\markboth{D. Indumathi and H. Saveetha}
{Study of vector meson fragmentation using a broken SU(3) model}

%%%%%%%%%%%%%%%%%%%%% Publisher's Area please ignore %%%%%%%%%%%%%%%
%
%\catchline{}{}{}{}{}
%
%%%%%%%%%%%%%%%%%%%%%%%%%%%%%%%%%%%%%%%%%%%%%%%%%%%%%%%%%%%%%%%%%%%%

\title{Study of vector meson fragmentation using a broken SU(3) model}

\author{D. Indumathi}
%\address{Institute of Mathematical Sciences, \\ CIT Campus, \\ Chennai 600 113, India \\ indu@imsc.res.in}
\affiliation{Institute of Mathematical Sciences, \\ CIT Campus, \\ Chennai 600 113, India \\ indu@imsc.res.in}

\author{H. Saveetha}

%\address{Department of Theoretical Physics, University of Madras, \\ Guindy Campus, Chennai 600 025, India. \\ saveehari@gmail.com}
\affiliation{Department of Theoretical Physics, University of Madras, \\ Guindy Campus, Chennai 600 025, India. \\ saveehari@gmail.com}

%\begin{history}
%\received{Day Month Year}
%\revised{Day Month Year}
%\end{history}

\begin{abstract}
Inclusive hadro production in $e^+\,e^-$ annihilation processes is
examined to study the fragmentation process. A broken SU(3) model
is used to determine the quark and gluon fragmentation functions of
octet vector mesons, $\rho$ and $K^*$, in a simple way with an SU(3)
breaking parameter $\lambda$. These are expressed in terms of just two
light quark fragmentation functions, $V(x, Q^2)$ and $\gamma(x, Q^2)$
and the gluon fragmentation function $D_g(x, Q^2)$. These functions
are parameterized at the low input scale of $Q_0^2 = 1.5$ GeV$^2$,
evolved through LO DGLAP evolution including charm and bottom flavour
at appropriate thresholds, and fitted by comparison with data at the
$Z$-pole. The model is extended with the introduction of a few additional
parameters to include a study of singlet--octet mixing and hence $\omega$
and $\phi$ fragmentation. The model gives good fits to the available
data for $x \gtrsim 0.01$, where $x$ is the scaled energy of the
hadron. The model is then applied successfully to $\omega$, $\phi$
production in $p\,p$ collisions at the Relativistic Heavy Ion Collider,
RHIC; these data form an important base-line for the study of Quark Gluon
Plasma in heavy nucleus collisions at RHIC, and also in future at the
LHC.


\end{abstract}

%\ccode{PACS numbers: 13.60.Le, 13.60.Hb, 13.66.Bc, 13.85.Ni}

\maketitle

\section{Introduction}
\label{Intro}
Studies of meson fragmentation are currently limited by data in comparison
with the relatively abundant data available on deep inelastic processes
(DIS). Meson fragmentation processes can be understood within QCD through
time-like conjugates of the space-like processes that contribute to
DIS. Hence there is a great deal of interest in the study of meson (and
baryon) fragmentation. Data from $e^+\,e^-$ collisions are most commonly
available for pseudo-scalar and vector meson fragmentation. While $e\,p$
data is severely limited, there has recently been high quality data on
pseudo-scalar meson production from $p\,p$ collision processes in
RHIC\cite{RHIC}. Preliminary data on light mesons is already available
from the LHC\cite{Dasu} and more is expected shortly. In particular,
an understanding of $\eta$ and $\phi$ meson fragmentation in $p\,p$
processes is important as a baseline for the study of the
production of these mesons in nucleus--nucleus collisions as
a signal of quark--gluon plasma (QGP)\cite{qgplhc}. While
there exist many phenomenological studies on $\pi$ and $K$
meson\cite{Indu,Misra,Albino1,Albino2,Albino3,Kumano1,Kumano2} and $\eta$
fragmentation\cite{Misra,Stratmann}, as well as many comprehensive
reviews\cite{Albino1,Albino2,Albino3,Ams} of these, the issue of vector
meson hadro-production has not been addressed so far. In this paper
we focus attention for the first time on light ($u$,$d$, $s$ valence
quarks only) vector meson fragmentation using a model that has earlier
been applied to a study of light pseudo-scalar meson\cite{Indu,Misra}
and octet baryon\cite{Indu} fragmentation. In particular, the $\phi$
meson, which is almost a pure strange quark--anti-quark bound state, has
special relevance as a signal for QGP\cite{RHICphi,RHICphipt1,RHICphipt2}.

A study of fragmentation functions requires an experimental input at
a given $Q^2$ (momentum transfer) scale since QCD cannot predict the
fragmentation functions themselves but only their $Q^2$ dependences.
In the case of mesons formed from the light quarks ($u$,
$d$, $s$), it is possible to apply symmetry arguments to reduce further
the number of (unknown) input starting fragmentation functions.

A simple SU(3) symmetric model is introduced\cite{Indu} which
has been applied to pseudo-scalar octet mesons and octet baryons at the 
leading order level\cite{Indu,Misra}. In this paper, the above model 
is applied to study the fragmentation functions of octet vector mesons 
($\rho$ and $K^*$) in $e^+\,e^-$ annihilation. The model reduces 
the various required input quark fragmentation functions to a 
combination of just three independent fragmentation functions, 
$\alpha(x, Q^2)$, $\beta(x, Q^2)$, $\gamma(x,Q^2)$ and an SU(3) 
breaking scale-independent parameter $\lambda$ at a low input scale. 
In addition to these functions, the gluon and heavy quark contributions 
are also taken into consideration during evolution. The model is then 
extended to predict the octet-singlet mixing of $\omega,\phi$ mesons with a 
very simple ansatz that relates the single fragmentation function in 
the singlet sector to an octet fragmentation function
along with a few additional constants. In contrast, other
studies\cite{Stratmann,Kumano1,Kumano2,Albino1,Albino2,Albino3} in, say,
the pseudo-scalar meson sector, fit individual data on different mesons
with no attempt made to combine the data in any way whatsoever. This
is in fact one possible reason why the vector meson sparse data set may
have not been studied so far.

The paper is organized as follows: the kinematics of the relevant
$e^+\,e^-$ and $p\,p$ scattering processes is briefly explained in section
$2$. The model to study the fragmentation functions of vector mesons is
introduced in section $3$. The model is used to study the cross-section
for hadro-production in section $4$. In particular, a detailed study of
the pure octet mesons, $\rho$ and $K^*$, is given in this section. It is
then extended to include the singlet case, with singlet--octet mixing,
in section $5$. The detailed parameterization is explained in section
$6$. Finally, the model is compared with both $e^+\,e^-$ and $p\,p$ data
in section $7$. A summary and discussion is presented in section $8$
of the paper.

\section{Formalism}
\label{Formalism}
\subsection{Hadron production in $e^+\,e^-$ process}
\label{Formalism-sigma}
The $e^+\,e^-$ annihilation process is used to analyze the fragmentation
functions of quarks (anti-quarks) through the hadrons which they
produce. The reaction $e^+\,e^- \rightarrow h+X~$ proceeds by first
creating a quark and an anti-quark pair through $e^+\,e^- \rightarrow q
\overline{q}$ via an intermediate vector boson $V=\gamma/Z^0$, followed
by the fragmentation of the quark$(q)$, anti-quark$(\overline{q})$, or
gluon $(g)$ into a hadron. This process is known as the fragmentation
process\cite{Hirai}.

\begin{figure}[pb]
%\centerline{\psfig{file=eeprocess.eps,width=0.7\textwidth}}
\includegraphics[width=0.7\textwidth]{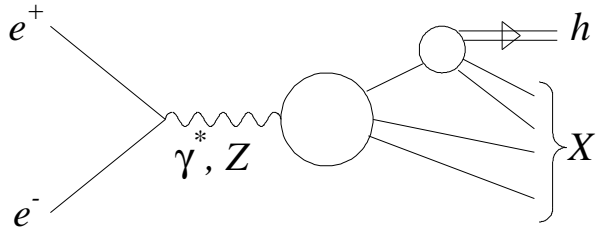}
\vspace*{8pt}
\caption{A schematic diagram of the fragmentation process in $e^+\,e^-$
scattering\protect\cite{Bin}.}
\label{eep}
\end{figure}

The term $h$ in Fig.~\ref{eep} represents the hadron (vector meson in our
case), where $X$ is the debris. The corresponding cross-section for this
scattering process at c.m. energy $\sqrt{s}$ factorises into a component
describing the hard scattering and one describing the hadronisation.
Hence, it can be expressed to LO as\cite{Ams}:
\begin{eqnarray}
\frac{1}{\sigma_{tot}}\frac{\d \sigma^h}{\d x} & = & \frac{\sum_q c_q \,
D_q^h(x, Q^2)}{\sum_q c_q}~,
\label{eq:cross}
\end{eqnarray}
in which the fragmentation function $D_q^h(x, Q^2)$ is the probability
for a quark to hadronise to a hadron carrying a fraction $x$ of the
energy from its parent quark, where $x \equiv E_{hadron}/E_{quark}
=(2E_{h}/\sqrt{s}) \leq1$ (or $x_p \equiv 2p_h/\sqrt{s}$) and Q
=$\sqrt{s}$ is the energy scale, where we are considering the reaction.

The charge factors $c_q$ are associated with the quark $q_i$ with flavor
$i$, written\cite{Ams} in terms of the electromagnetic charge $e_i$, vector
and axial vector electroweak couplings, $v_i=T_{3i}-2e_i\sin^2\theta_{\rm
w}$ and $a_i=T_{3i}$, as
\begin{eqnarray} \nonumber
c_q & = & c_q^V + c_q^A~, \\ \nonumber
c_q^V & = & \frac{4 \pi \alpha^2}{s}[e_q^2 + 2e_q v_e v_q \, \rho_1(s)
+(v_e^2 + a_e^2)v_q^2 \, \rho_2(s)]~, \\ \nonumber
c_q^A & = & \frac{4 \pi \alpha^2}{s}(v_e^2+a_e^2)\ a_q^2\ \rho_2(s)~, \\
\rho_1(s) & = & \frac{1}{4 \sin^2 \theta_{\rm w} \cos^2 \theta_{\rm w}}
 \ \frac{s(m_Z^2-s)}{(m_Z^2-s)^2+m_Z^2 \Gamma_Z^2}~, \\ \nonumber 
\rho_2(s) & = & \frac{1}{(4 \sin^2 \theta_{\rm w} \cos^2 \theta_{\rm
w})^2} \ \frac{s^2}{(m_Z^2-s)^2+m_Z^2 \Gamma_Z^2}~. \nonumber
\end{eqnarray}
The values of $T_{3i}$, the third component of weak isospin, and $v_i$
are tabulated  with weak mixing angle $\theta_{\rm w}$ in\cite{Hal}.
$\Gamma_Z$ and $m_{\rm Z}$ are the decay width and mass of the
$Z$-intermediate gauge boson for high energy scale. We re-express the
LO cross section in terms of singlet and non-singlet fragmentation
functions, as
\begin{eqnarray}
\frac{1}{\sigma_{tot}} \frac{\d \sigma^h}{\d x} & = & \frac
{\sum_j {a_j D_j(x,t)}}{\sum_q {c_q}} ~,
%{a_0 D_0(x_E,t) +a_3 D_3(x_E,t) + a_8 D_8(x_E,t)}{\sum_q {c_q}} ~,
\label{eq:sigtot}
\end{eqnarray} 
where $j = 0, 3, 8, 15, 24$ and $D_0$, $D_3$, $D_8$, $D_{15}$ and $D_{24}$
refer to the singlet ($D_0 = \sum_i (D_i + D_{\overline{i}})$), and
the non-singlet ($n^2-1=3, 8, 15, 24$: $(u - d)$, $(u + d - 2 s)$,
$(u+d+s-3c)$ and $(u+d+s+c-4b); u \equiv D_u + D_{\overline{u}},$
etc.)  combinations respectively. Here the coefficients are
$a_0 = (c_u+c_d+c_s+c_c+c_b)/5$, $a_3 = (c_u-c_d)/2$, $a_8 =
(c_u+c_d-2c_s)/6$, $a_{15} = (c_u+c_d+c_s-3c_c)/12$ and $a_{24} =
(c_u+c_d+c_s+c_c-4c_b)/20$. Note that the data from LEP or SLD on the
$Z$-pole is dominated by the $D_0$ fragmentation function combination
while at lower energies where photon exchange dominates, the
$u$-type quark fragmentation functions dominate. Furthermore,
data is available\cite{RhoZ1,RhoZ2,RhoZ3,RhoHRS} only
for the sum of the conjugate states, $\rho^\pm = \rho^+
+ \rho^-$, and similarly for the charged and neutral $K^*$
mesons\cite{KstarALEPH1,KstarALEPH2,KstarDELPHI,KstarOPAL}. We therefore
compare with the sums of these quantities. The SLD data\cite{SLD1,SLD2}
is available separately for fragmentation from light quarks ($u,d,s$)
alone; hence for these data, the summation in Eq.~(\ref{eq:cross}) is over
the light flavour contributions only. Data on the singlet--octet mixture
of states, viz., $\omega$ and $\phi$, are available from both LEP and
SLD\cite{Omega1,Omega2,PhiALEPH1,PhiALEPH2,PhiDELPHI,PhiOPAL,SLD1,SLD2}.

\subsection{ Hadron production in $p\,p$ process}
\label{Formalism-sigmapp}

In addition to information on the unknown fragmentation functions,
hadro-production in $p\,p$ processes requires a knowledge of the parton
(both quark and gluon) distribution functions within the primary
proton. The scattering is expressed in terms of the underlying parton
interactions, with one of the final state partons fragmenting into the
hadron of interest. All possible initial state parton interactions are
possible: $q\,q$, $q\,g$ and $g\,g$, as well as processes with antiquarks.

The inclusive cross section for hadro-production, $p+p \rightarrow
h+X$, producing a hadron $h$ at large $p_T$ is given
by\cite{Reyafrag,OK,BBK},
\begin{equation}
E_h\frac{\d^3\sigma}{\d p^3_h} = \frac{1}{\pi} \sum \int_{x_a^{\rm min}}^1
           \d x_a \int_{x_b^{\rm min}}^1 \d x_b\,P_a^A(x_a,Q^2)\,P_b^B(x_b,Q^2)
	   \,\frac{\d\sigma^{ab \rightarrow cd}}{z_h \d \hat{t}} \,
	   D_c^h(z_h,Q^2)~,
\label{eq:pp}
\end{equation}
where the sum over $(a,b,c,d)$ runs over both quarks and gluons. Here
$x_a$ and $x_b$ are the usual Bjorken-$x$ variables corresponding to the parent
proton momenta $p_A$ and $p_B$: $x_a=p_a/p_A$, $x_b=p_b/p_B$ (neglecting
intrinsic transverse momentum) and
$P_{a/A}(x_a, Q^2)$ are the usual parton density distributions; for
example, $P_{u/p}(x_a, Q^2) \equiv u(x_a, Q^2)$, etc.

The fragmentation functions depend on the variables, $z = z_h =
p_h/p_c$, the fraction of momentum of the quark carried by the
fragmenting hadron, and the scale
$Q^2 \sim p_T^2$.
The limits of integration are\cite{Reyafrag,OK,BBK}
\begin{equation}
\label{eq:xmin}
x_a^{min} = \frac{x_1}{1-x_2}~;~~
x_b^{min} = \frac{x_a x_2}{x_a-x_1}~, \nonumber
\end{equation}
with $x_1=-u/s, x_2=-t/s$.

For numerical comparison with the data, we re\"express the
cross-section in terms of the physical observables which are the
transverse momentum $p_T = p_h \sin\theta$ and the rapidity
$y=(1/2)\ln [(E_h + p_h \cos\theta)/(E_h - p_h \cos\theta)]$, as
\begin{equation}
E_h\frac{\d^3\sigma}{\d p^3_h} \equiv
\frac{1}{p_T}\frac{\d^3\sigma}{\d p_T \d y \d\phi}~,
\label{eq:pTy}
\end{equation}
where $\theta$ is the scattering angle of the hadron $h$ in the
$p\,p$ center of mass frame and $E_h$ and $p_h$ are its energy and
3-momentum. Note that the azimuthal angle dependence is trivial in
this process; furthermore the data from the PHENIX experiment at
RHIC is taken over a range $\pi$ in $\phi$ and $-0.35 \le y \le 0.35$
in the rapidity\cite{RHICphi,RHICphipt1,RHICphipt2}. The sub-process
cross-sections are well-known\cite{Reyafrag,OK,BBK,CM}; the $q\,q$, $q\,g$
and $g\,g$ processes all contribute at the same order in $\alpha_s$. Hence
the quark and gluon fragmentation functions contribute at the same order,
unlike in the $e^+\,e^-$ case. This data is therefore an important test
of the correctness of the gluon fragmentation functions.

We now present details of our model for quark fragmentation functions.

\section{SU(3) Model}
\label{model}
Fragmentation functions parameterize the hadronisation process, in
which the observed hadrons are formed from the final state partons of
the scattering process; these cannot be calculated in QCD.  However,
given a definite energy scale and starting distribution, QCD can evolve
these perturbatively and explain their scale $(Q^2)$ dependence.  A set
of common fragmentation functions are used to describe the members of
the octet of vector mesons $\rho(\rho^+, \rho^-, \rho^0)$, $K^*(K^{*+},
K^{*-}, K^{*0}, \overline{K}^{*0})$ and $\omega$. An SU(3) symmetric
model (with parameterized SU(3) breaking) has been chosen to achieve this
because, in principle, such a symmetry gives good description about the
octet of vector mesons. Such a model with broken SU(3) was developed
for pseudo-scalar octet $\pi$ and $K$ meson fragmentation in $e^+\,e^-$
collisions\cite{Misra} and for the octet baryons\cite{Indu} $p,n, \Lambda,
\Sigma$. We use an analogous model in the vector meson sector in
our analysis.

We start with light quarks ($u$, $d$ and $s$) at the starting scale of
$Q_0^2 = 1.5$ GeV$^2$. However, as the fragmentation function evolves
across various thresholds (typically up to $Q^2 = (91.2$ GeV)$^2$), consistent
contribution of charm and bottom quarks are included appropriately in
the evolution. Let us consider the process at the input scale as,
$$
q_i \rightarrow h^i_j + X_j~.
$$
The underlying SU(3) process can be thought of as $3 \rightarrow 8
+ X$.  That is, a quark goes to an octet hadron $(h^i_j)$ with the
remainder $X_j$ being a triplet $(3)$, antisixplet $(\overline{6})$ or
fifteenplet $(15)$ with $i,j$ running over 1 to 8. Let $\alpha(x,Q^2)$,
$\beta(x,Q^2)$ and $\gamma(x,Q^2)$ be the corresponding unknown SU(3)
symmetric independent fragmentation functions for each of these
possibilities\cite{Indu}, that is, for X to be 3 $(\overline{6},15)$
the probability of the quark to fragment into an octet meson is $\alpha$
($\beta, \gamma)$.

In a similar way, an anti-quark also produces an octet hadron
with $X_j$ being an anti-triplet ($\overline{3}$), sixplet (6) or
anti-fifteenplet ($\overline{15}$), for which $\overline{\alpha}(x,Q^2)$,
$\overline{\beta}(x,Q^2)$ and $\overline{\gamma}(x,Q^2)$ have to
be determined.Thus a single meson has seven unknown fragmentation
functions $D_q^h(x, Q^2)$, ${\overline{D}}_{\overline{q}}^h(x, Q^2)$
and $D_g^h(x, Q^2)$ associated with its production. Here $D_q^h$,
${\overline{D}}_{\overline{q}}^h$ and $D_g^h$ refer to the light quark,
anti-quark and gluon fragmentation functions while the heavier quark
contributions are zero at the starting scale (below the charm threshold).

So, we have to fit a total of 56 $(8 \times 7)$ unknown
fragmentation functions to the data for octet mesons, which is
rather daunting. The problem is made simpler when we apply SU(3)
symmetry since the seven independent symmetric fragmentation functions
$\alpha(x,Q^2)$, $\beta(x,Q^2)$ and $\gamma(x,Q^2)$ including their
conjugates $\overline{\alpha}(x,Q^2)$, $\overline{\beta}(x,Q^2)$ and
$\overline{\gamma}(x,Q^2)$ and $D_g(x, Q^2)$, the gluon contribution,
should determine the fragmentation of the entire group of octet
mesons. Since SU(3) symmetry is only approximate, we use a single
$x$-independent parameter to signify SU(3) breaking, while SU(2) remains
unbroken in our model. Hence the fragmentation functions of $\rho^\pm$
and $\rho^0$ (and similarly for isospin conjugates of $K^*$) are related
by isospin symmetry.

\subsection{Valence and Sea functions}
\label{model-valsea}
The vector meson octet is a self conjugate octet. So, $D^h_q =
{\overline{D}}^h_{\overline{q}}$~. Therefore, we have three independent
fragmentation functions as mentioned in the above section and we express
the quark fragmentation in terms of these three fragmentation functions in
Table~\ref{tab:frag}. We reduce the number of unknown functions further
through various symmetry considerations like \emph{isospin invariance
and charge conjugation}. We assume the sea is flavour symmetric, so
that $D_u^{\rho-} = D_s^{\rho-}$ and so on.  Using this assumption
and the expressions for corresponding fragmentation functions given in
Table~\ref{tab:frag}, we have
\begin{equation}
\beta(x, Q^2)  = \gamma(x, Q^2) /2~;
\label{eq:beta}
\end{equation}
and all the sea fragmentation functions are equal to
\begin{equation}
S(x,Q^2) = 2 \gamma(x,Q^2)~.
\label{eq:sea}
\end{equation}
Thus, all valence fragmentation functions can be expressed in terms of
the function $V(x, Q^2)$, where $V$ is given, say for $\rho^+$, by the
difference $D_u^{\rho+}- D^{\rho+}_{\overline{u}}$. Therefore, we have
\begin{equation}
V(x,Q^2) = \left(\alpha + \beta + \frac{3}{4}\gamma\right)(x,Q^2) 
- 2\gamma(x,Q^2)~,
\label{eq:valence}
\end{equation} 
Substituting the value of $\beta$ from Eq.~(\ref{eq:beta}) in 
Eq.~(\ref{eq:valence}) we get the valence combination to be
\begin{equation} 
V(x, Q^2) = \alpha(x,Q^2) - \frac{3}{4}\gamma(x,Q^2)~.
\label{eq:val}
\end{equation} 
All sea fragmentation functions can be expressed in terms of
$\gamma$. Thus, Eqs.~(\ref{eq:sea}) and (\ref{eq:val}) represent just
two unknown fragmentation functions for the sea $S$ (or equivalently 
$\gamma$) and valence $V$, in terms of which all quark fragmentation 
functions which describe all the octet vector meson production 
can  be represented.

\subsection{Breaking of SU(3) symmetry}
\label{model-breaking}
SU(3) symmetry is broken in the model due to relatively more massive
strange quarks.  So, in addition to valence and sea functions, we
introduce an $x$-independent symmetry breaking parameter $\lambda$
for a non-strange quark to fragment into a strange octet meson.

For example, to produce $K^{*+}$ ($u\overline{s}$) meson, a $u$ quark
in the valence has to pick up a(n anti) $s$ quark; being more massive,
the corresponding fragmentation function is suppressed by the parameter
$\lambda$.  Moreover, if the valence quark is $\overline{s}$ then it
may easily pick up a $u$ quark without this suppression factor. In
the same way, fragmentation of other mesons like $K^{*-}$, $ K^{*0}$
and ${\overline{K}}^{*0}$ can be explained.

For all these mesons with strange quark in their valence, apart
from different valence parts, their sea function remains the
same---$2\lambda\gamma$---because the sea is flavour symmetric and is
uniformly suppressed by the suppression factor $\lambda$. Thus we use
\emph{broken SU(3) symmetry} with an SU(3) symmetric sea as our model
to describe the fragmentation functions of vector mesons with the
introduction of the parameter $\lambda$.

\section{Fragmentation of $\rho$ and $K^*$ mesons} 
\label{rhokstar}
We begin by neglecting the problematical $\omega$ meson that is not
a pure SU(3) octet meson; it will be considered along with the $\phi$
meson later.

\subsection{ $\rho$ meson}
\label{subrho}
The scattering cross section is expressed in terms of singlet and
non singlet combinations in Eq.~(\ref{eq:sigtot}).  Let us begin with
$\rho$, the lightest meson, and express the fragmentation function for
$u$ quark from Table~\ref{tab:frag}. On substituting the values of sea
and valence parts from Eqs.~(\ref{eq:sea}) and (\ref{eq:val}) we get,
$$
D_u^{\rho^{+}} = V + 2\gamma~,
$$
as $u$ quark is present both in the valence as well as in the sea part 
and is equal to the $\overline{d}$ contribution. The charge conjugation
invariance clearly shows that the other quarks are not in the valence
of the $\rho$ meson; hence their fragmentation for sea part has to
be $2\gamma$. The singlet contribution at the input scale for
$\rho$ meson (where only the three light flavours contribute) is
therefore the total quark contribution:
\begin{eqnarray}
D_0^{{\rho}^+} =\ D_{u+{\overline{u}}+d+{\overline{d}}+
                    s+{\overline{s}}}^{{\rho}^+} \quad
               =\ 2V + 12\gamma~.
\label{eq:D0rho}
\end{eqnarray}      
The non singlet contributions $D_3$ and $D_8$ can be obtained for 
$\rho$ meson with the same analogy. Here, the non singlet 
term $D_3$ ($= D_u-D_d$) turns out to be zero due to charge conjugation 
invariance and $D_8 = 2V$.
\subsection{$K^*$ meson}
\label{subkstar}
The procedure for the $K^*$ meson is exactly the same as above. The
only difference is that it has a strangeness quantum number. Thus
for the valence component of $K^{*-}$ meson ($s\overline{u}$), the
non-strange $u$ quark contribution is suppressed by $\lambda$ as it has
to pick up a strange quark to form the $K^{*-}$ meson. However, the ${s}$
contribution is itself not suppressed, since only an (anti-)$u$ quark is
required here. Meanwhile, all the sea fragmentation functions come with a
uniform factor of $\lambda$ since for all quarks (flavour symmetric sea),
a strange quark has to be produced. Therefore,
\begin{equation}
D_0^{K^{*+}}  =\  D_{u+{\overline{u}}+d+{\overline{d}}+
                    s+{\overline{s}}}^{K^{*+}} \quad
              =\  (1 + \lambda)V + 12\lambda\gamma~.
\label{eq:D0Kstar}
\end{equation}
Notice that while $\lambda$ is $x$-independent, there is an inherent
$x$-dependence of the strangeness suppression, being maximal at small-$x$
and least at large-$x$ due to the different suppression of the valence
and the sea quarks.

Thus $\rho$ and $K^*$ mesons have different valence and sea 
quark fragmentation functions. In addition, since the gluon 
fragmentation function mixes with the singlet $D_0$ fragmentation 
function on evolution, we parameterize a possible gluon suppression 
through $D^{K^*}_g = f^{K^*}_g D^{\rho}_g$. 

\section{Extension to $\omega$ and $\phi$ mesons}
\label{extnu1}
We now extend this \emph{broken SU(3) model} which explains in a most
simple way the pure octet ($\rho$ and $K^*$) mesons, to the  $\omega$
and $\phi$ mesons, which are orthogonal combinations of the SU(3) octet
($\omega_8$) and singlet states ($\omega_1$):
\begin{eqnarray} \nonumber
\omega & = & \sin\theta~\omega_8 + \cos\theta~\omega_1~, \\
\phi   & = & \cos\theta~\omega_8 - \sin\theta~\omega_1~.
\label{eq:statedef}
\end{eqnarray} 
where $\omega_8 = (u\overline{u} + d\overline{d} -
2s\overline{s})/\sqrt{6}$, $\omega_1 = (u\overline{u} + d\overline{d}
+ s\overline{s})/\sqrt{3}$ are the corresponding orthogonal states
and $\theta$ is the vector mixing angle, whose value is
approximately\cite{PDG} $35^{\circ}$. Note that a value of $\theta$
close to this value saturates the physical $\phi$ state\cite{Yao} as a pure
$s\overline{s}$ state. The fragmentation
functions for $\omega_8$ can be described with the help of fragmentation
functions given in Table~\ref{tab:frag}, since $\omega_8$ is one of
the members of the octet, whereas details regarding $\omega_1$ will be
discussed in the next section.

\subsection{Singlet hadron ($\omega_1$) fragmentation}
\label{o1}
Let us consider the same process,
$$
q_i \rightarrow h + X_i~,
$$
in which a quark hadronises into a singlet meson so that $X$ can only
be a triplet ($3 \rightarrow 1 + X$). Therefore, we need to determine
only one unknown fragmentation function $\delta(x, Q^2)$ in the singlet
case. In section 3 we saw that the probability for a parton to fragment
into an octet hadron with $X$ being triplet is $\alpha(x, Q^2)$. For
fragmenting to a meson so that the remainder $X$ is a triplet state,
therefore, there are only two possibilities: either the hadron is an octet
(the process is proportional to $\alpha$) or the hadron is a singlet
(the process is proportional to $\delta$). Hence we use the simple
ansatz that the function $\delta$ is simply related to $\alpha(x,
Q^2)$, the fragmentation function for members of octet meson. That is,
\begin{equation}
\frac{\delta}{3} = \frac{f_1 \,\alpha}{3} = \frac{f_1}{3}\left(V +
\frac{3}{4}\gamma\right)~,
\label{eq:delta}
\end{equation}       
where the factor of $1/3$ is due to the normalization of the state and
$f_1$ is the (presumed) $x$-independent proportionality constant we have
to determine in the analysis. Since this uses an approximate SU(3)
symmetry, this constraint is applied only at the input scale $Q_0^2$
where there are only three active flavours. The evolution is correctly
applied to all the active flavours, depending on the scale.

We will now express the fragmentation of $\omega$ and $\phi$ mesons in
terms of the SU(3) octet and singlet contributions.

\subsection{Singlet and octet contribution to $\omega$ and $\phi$ mesons}
\label{omphi}

The strangeness suppression factor $\lambda$ remains the same
here. However, there is a difference in the sea suppression factor. While
it was also equal to $\lambda$ in the case of the $K^*$ meson, here the
suppression is rather different. It arises due to the preference of a
given $q\overline{q}$ pair to fragment to the lighter $\rho$ rather
than to $\omega$ or $\phi$. In particular, in the case of $\omega$,
which is mostly saturated by non-strange quarks in the valence, the
sea suppression factor arises due to the preference of a given light
$q\overline{q}$ pair to fragment to the lighter $\rho$ rather than the
$\omega$. Hence the sea suppression factor is expected to be $f_{\rm
sea}^\omega \sim  m_\rho^2 / m_\omega^2$ just as in the case of
pseudoscalar mesons\cite{Misra}, and we
do not expect a large suppression. On the other hand, the physical $\phi$
state is saturated by the strange contribution; hence the suppression
factor $f_{\rm sea}^\phi$ for $\phi$ is expected to be close to
$\lambda^2$ as a quark has to pick up both s and $\overline{s}$ from 
flavour symmetric sea. We will see in the next section that numerical 
fits to the data do indeed agree with these expectations.

We first explicitly write down the fragmentation functions given in
Table~\ref{tab:frag} for corresponding octet hadrons in terms of valence
and sea sectors, including the various suppression factors as discussed
above:
\begin{eqnarray}
D_u^8 &=& \frac{V}{6} + 2 f_{\rm sea} \gamma~, \\ \nonumber
D_s^8 &=& \frac{2}{3} \lambda V + 2 f_{\rm sea}\gamma~;
\label{du8}
\end{eqnarray}
where the sea $\gamma$ and valence $V$ have their usual definitions,
described in Eqs.~(\ref{eq:sea}) and (\ref{eq:val}), $\lambda$ is the
strangeness suppression factor while $f_{\rm sea}$ is the unknown
suppression factor for the SU(3)-symmetric sea fragmentation functions.

Using our ansatz for the singlet hadron, we have
\begin{eqnarray}
D_u^1  = D_d^1 &=& \frac{f_1^u}{3} \left( V + 
\frac{3}{4} f_{\rm sea} \gamma \right)~, \\ \nonumber
D_s^1  &=& \frac{f_1^s}{3} \left( \lambda V + 
\frac{3}{4} f_{\rm sea} \gamma \right)~.
\label{du1}
\end{eqnarray}
Here we have introduced separate suppression factors for the $u, d$-
and $s$-type singlet fragmentation functions. With these four equations
in hand, we express the fragmentation functions for $\omega$ and $\phi$
mesons, given the definitions of the states in Eq.~(\ref{eq:statedef}),
with vector mixing angle $\theta$, at the input scale as,
\begin{eqnarray} 
D_i^\phi &=& (c_i^\phi)^2\left(\cos^2\theta \frac{D_i^8}{(c_i^8)^2} +
\sin^2\theta \frac{D_i^1}{(c_i^1)^2}\right); \\ \nonumber
D_i^\omega &=& (c_i^\omega)^2\left(\sin^2\theta \frac{D_i^8}{(c_i^8)^2} +
\cos^2\theta \frac{D_i^1}{(c_i^1)^2}\right)~.
\label{duop}
\end{eqnarray}
Here, $i$ refers to the three light quarks ($u$, $d$, $s$);
the co-efficients are $c_u^\phi = c_d^\phi = (\cos\theta -
\sqrt{2}\sin\theta)$, $c_s^\phi = (-2\cos\theta - \sqrt{2}\sin\theta)$
and $c_u^8 = 1$, $c_s^8 = 2$, $c_u^1 = c_s^1 = \sqrt{2}$. Obviously,
we can find the coefficients for $\omega$ meson in the same way\cite{Misra}.
These fragmentation functions can be re-expressed in terms
of $D_0^\omega (x, Q^2)$, etc., in the usual way. Finally, we again 
parameterize the gluon fragmentation functions as $D^{\omega,\phi}_g 
= f^{\omega,\phi}_g D^{\rho}_g$.

\section{Parameterization of the input fragmentation functions}
\label{paramet}
The unknown functions for the valence $V(x, Q^2)$, sea  $\gamma (x,
Q^2)$, and gluon $D_g(x, Q^2)$ fragmentation are parameterized at low
input scale of $Q_0^2 = 1.5$ GeV$^2$ for three light quarks ($u$, $d$,
$s$) where the charm and bottom contributions $(D_c, D_b)$ are zero. The
parameters are then determined through fits with data. We use a standard
functional form to describe these quantities:
\begin{equation}
F_i(x) = a_i  x^{b_i}(1-x)^{c_i}(1 + d_i x + e_i x^2)~,
\label{eq:func}
\end{equation}
where $a_i$, $b_i$, $c_i$, $d_i$ and $e_i$ are the values to be determined.
The fragmentation functions are evolved to leading order (LO) including
charm and bottom contributions in appropriate places along with the
gluon fragmentation, to the $Q^2$ values of interest. Since the gluon
fragmentation does not occur in the expression for the 
cross-section, it is least constrained by these fits and relatively
unknown. Hence we set $e_g=0$ for the gluon.

\section{Data analysis and results}
\label{analysis}

\subsection{Fragmentation in $e^+\,e^-$ process}
We therefore have a set of parameterizations as well as a bunch of
scale-independent constants that need to be determined. We focus
here entirely on the $e^+\,e^-$ data as the cleanest sample. Hence
necessarily the fits to the gluons are ill-determined, since, to LO,
the gluon contributes only through evolution. Furthermore, we
concentrate on the $Z$-pole data\cite{Data} from both LEP as well
as SLAC-SLD. At this scale, five quark ($q\overline{q}$) flavour pairs
from $u$ to $b$ are produced in the final state and hence contributions
to the final state meson are of two types: first, is the direct
fragmentation of the quark (anti-quark) into a vector meson; another
is the fragmentation of a heavy (for example $D$ or $B$) meson which
subsequently decays into one of the mesons of interest. Since $b$ quarks
decay dominantly (greater than 95\% of the time) into $c$-quarks, the
decay chain of both heavy $b$ and $c$ end up in strange mesons, driven
by the large $V_{cs}$ CKM matrix element. We therefore expect that data
on the strangeness containing $K^*$ and $\phi$ mesons will have large
contamination from these heavy flavour decays.

On the other hand, the $K$ mesons decay predominantly into $\pi$ and
$\eta$; hence there is very little contamination of the $\rho$ and
$\omega$ data. In this analysis, therefore, we consider the inclusive
hadro-production data from LEP on the predominantly non-strange vector
mesons, $\rho$ and $\omega$\cite{Yao} to arise from light
quark fragmentation. For the strange mesons, $K^*$ and $\phi$, we use
instead the so-called $uds$ data from SLD where the fragmentation of
the light quarks alone have been segregated and studied.

\subsubsection{For pure octet mesons, $\rho$ and $K^*$}
\label{analysis-octet}

For extraction of fragmentation functions from the data we used inclusive
particle production in $e^+\,e^-$ as the process of choice. The
unknown fragmentation functions $V(x, Q^2)$, $\gamma(x, Q^2)$, $D_g(x,
Q^2)$ and the suppression factor $\lambda$ are determined by comparison
with the data\cite{Data}.

The comparison with data is restricted to the range $x \gtrsim 0.01$ since
it is well-known that the DGLAP evolution does not explain the behaviour
of the cross-section at small-$x$ and modifications such as MLLA (modified
leading log approximation) are needed to explain the small-$x$ data.

Note that individual data sets are some-what small, with typically 6--8
$x$-bins. This is insufficient to fit all the unknown fragmentation
functions (quark and gluon). However, the model can be applied to the
entire octet meson data as a whole; this allows for a reasonable fit to be
obtained with available data. This, in fact, is the primary motivation
for constructing such a model.

Using the available data we fit the functional form of Eq.~(\ref{eq:func})
for pure octet mesons ($\rho$ and $K^*$) by fixing the values of
parameters $a$, $b$, $c$, $d$ and $e$ for valence, sea and gluon
fragmentation functions.

The smaller $x$ behaviour is dominated by the sea ($\gamma$) contribution
and large $x$ behaviour by the valence $V$ contribution; this helps
determine the values of $a$,$b$, $c$. This process is not very sensitive
to the gluon contribution since this contributes only in the evolution
and not directly in the definition of the cross-section at LO, unlike in,
say, the $p\,p$ process.

\begin{figure}[pb]
%\centerline{\psfig{file=ijmpaf1.eps,width=4.7cm}}
\includegraphics[angle=-90, width=0.47\textwidth]{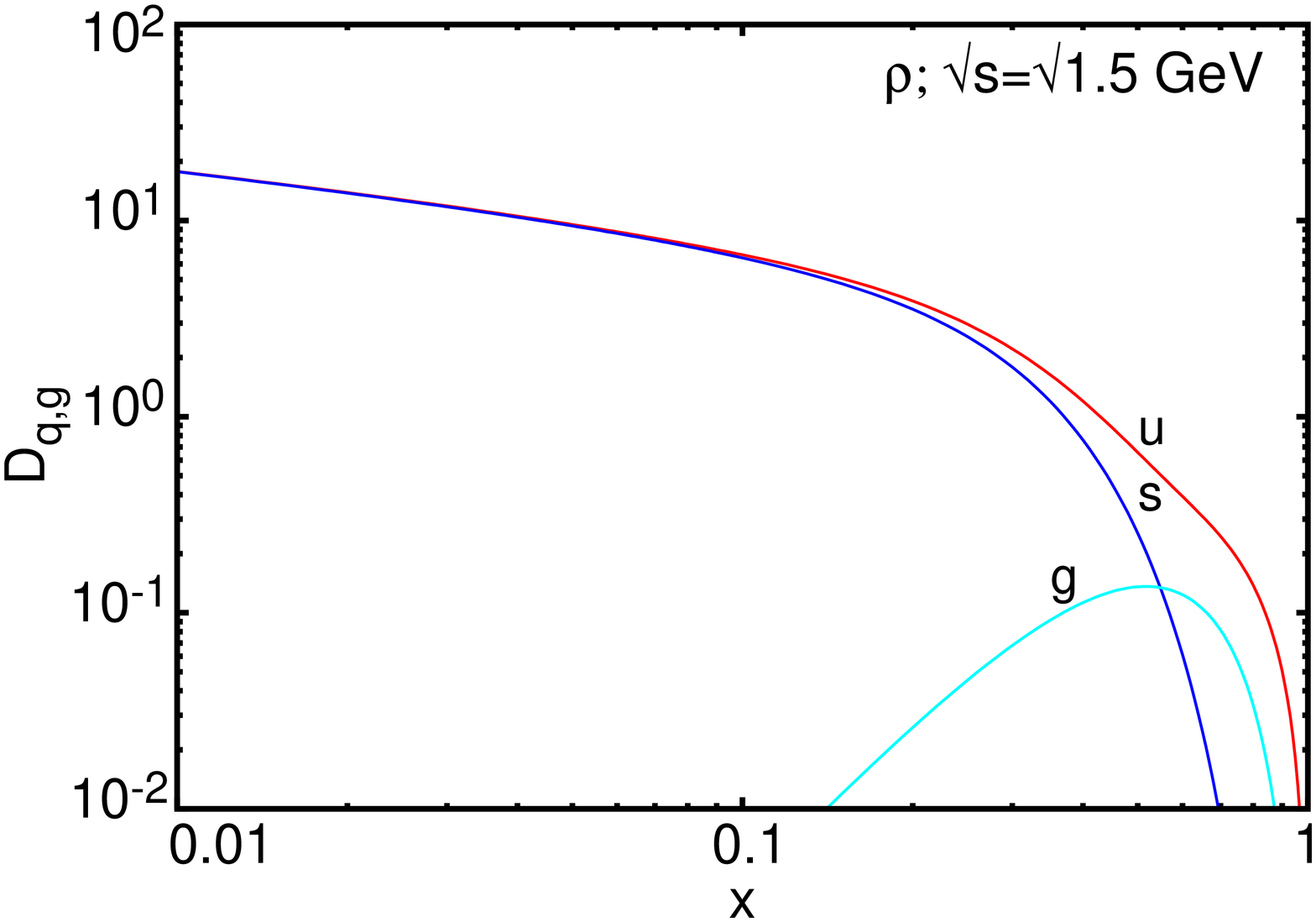}
\includegraphics[angle=-90, width=0.47\textwidth]{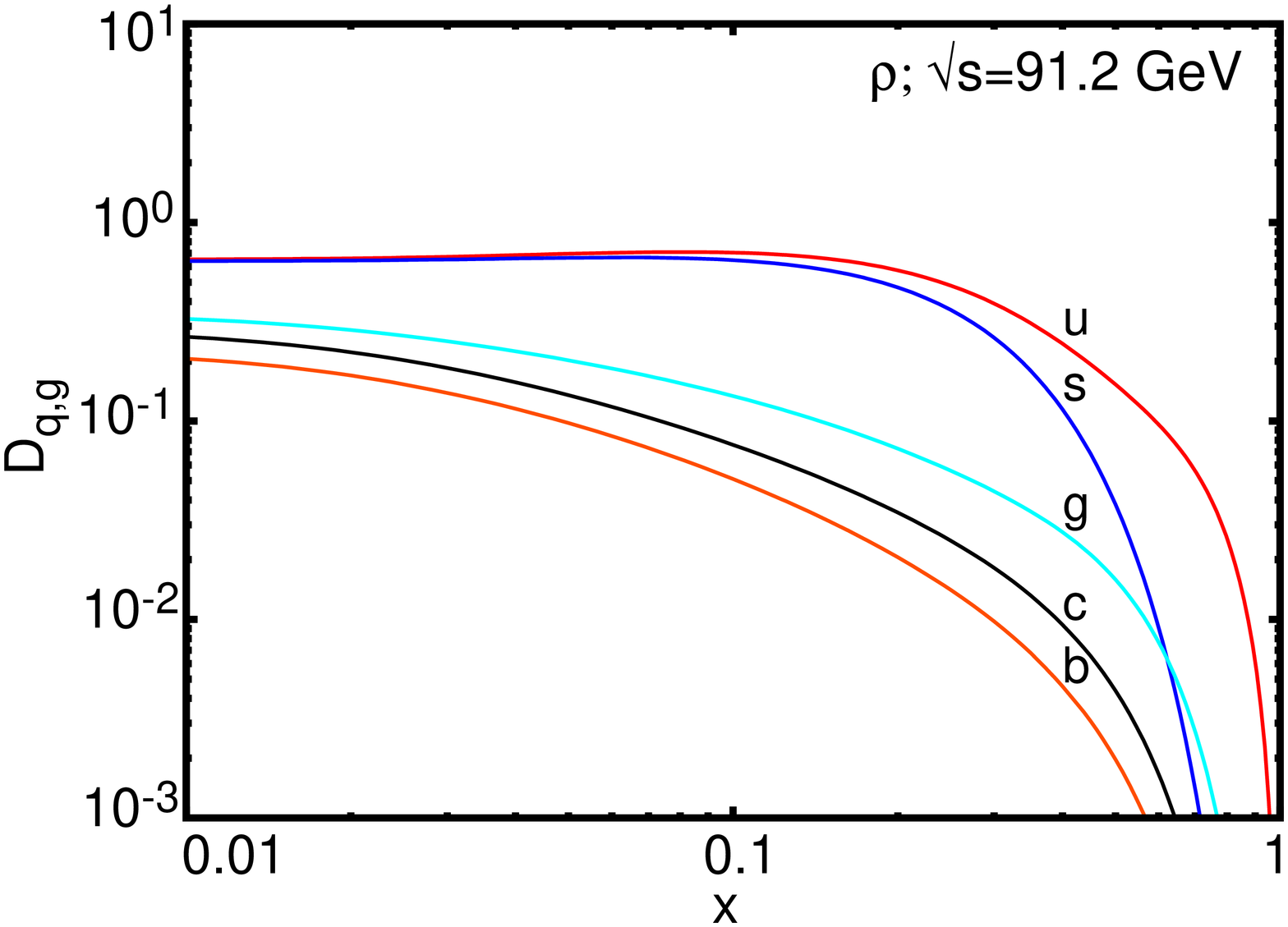}
\vspace*{8pt}
\caption{(L) Three input flavours $u=d$ and $s$ with gluon $g$ at the
starting scale of $\sqrt{s} = \sqrt{1.5} \ GeV$, as a function of $x$,
with zero contribution of $c$ and $b$ flavours for $\rho$ meson. (R)
Contribution of all the five flavours $u=d$, $s$, $c$ and $b$
with $g$ after leading order evolution at a scale of $\sqrt{s} = 91.2 \
GeV$ for $\rho$ meson.}
\label{fig:inputflav}
\end{figure}

\paragraph{The fragmentation functions}:
The best-fit values of the parameters in the input fragmentation
functions and the 1-$\sigma$ errors on them
are tabulated in Table~\ref{tab:inputs}. These correspond to the input
fragmentation functions for $\rho$ mesons having three light quarks
(the heavier $c$, $b$ quarks do not contribute) at the starting scale of
$Q_0^2 = 1.5$ GeV$^2$, as shown in Fig.~\ref{fig:inputflav}. This 
figure clearly shows that the valence contribution at large $x$ is 
dominated by $u$ and $d$ flavours, while strange flavour and gluon 
contribute only in the small $x$ sea part. As the evolution crosses 
various thresholds the charm and bottom contributions are included
appropriately as depicted in Fig.~\ref{fig:inputflav} which clearly 
reflects the (small) charm and bottom contribution after evolution at 
the $Z$-pole. The heavy flavour contribution purely arises from 
gluon-initiated processes and {\em cannot} account for the 
contribution in inclusive hadro-production data
arising from production and decay of heavy flavour mesons.

\begin{figure}[pb]
%\centerline{\psfig{file=ijmpaf1.eps,width=4.7cm}}
\includegraphics[angle=-90, width=0.49\textwidth]{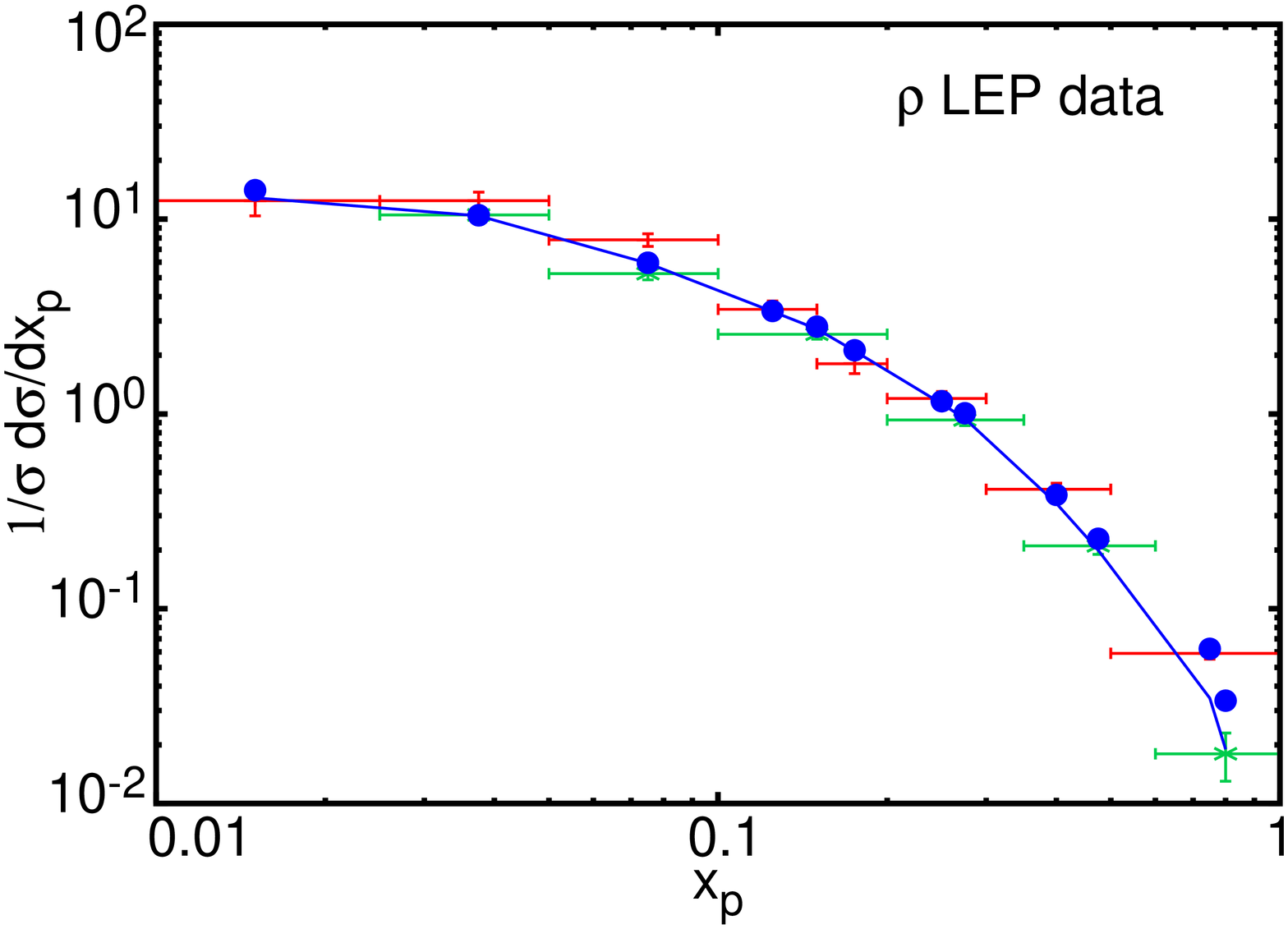}
\includegraphics[angle=-90, width=0.49\textwidth]{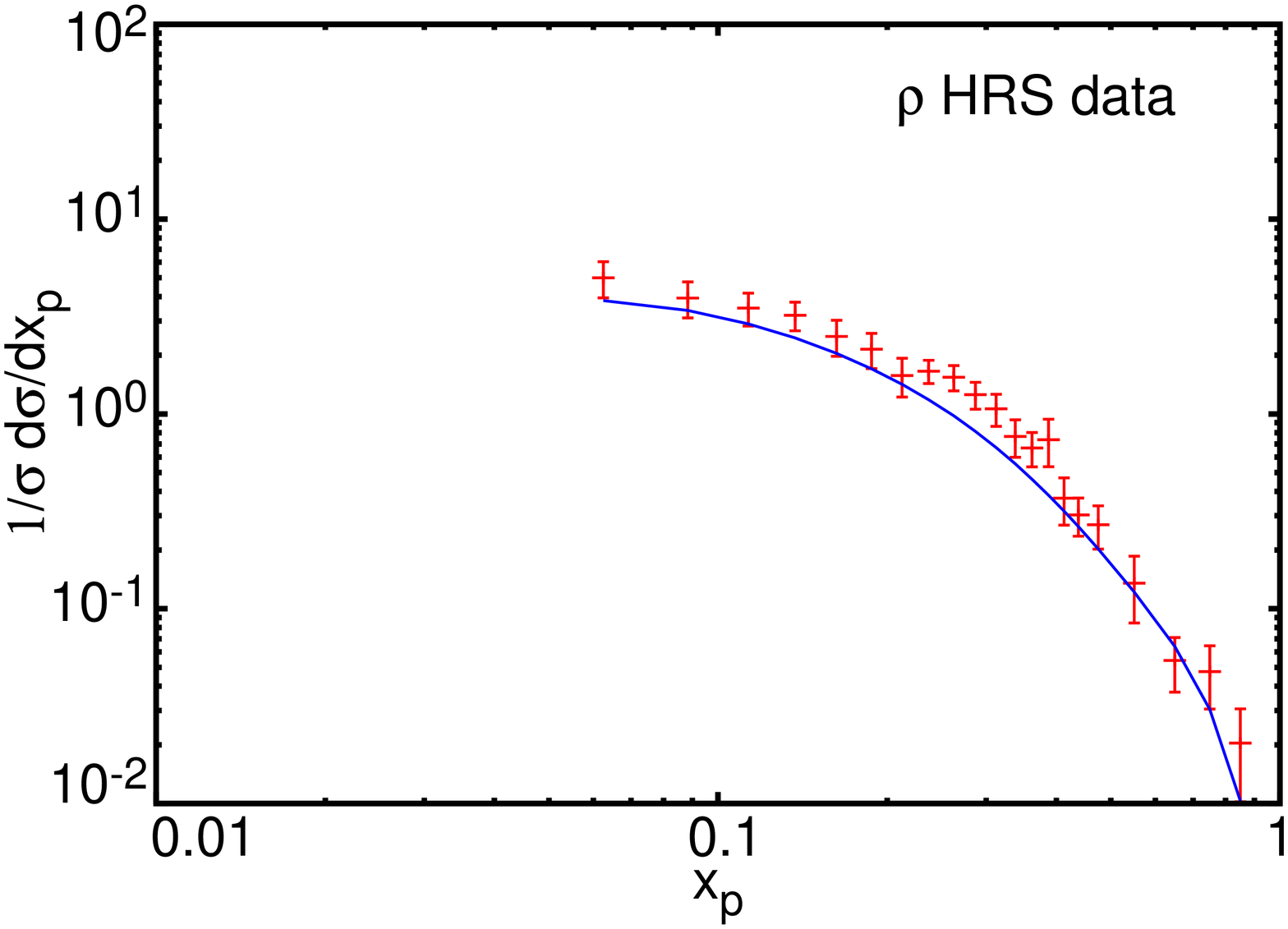}
\vspace*{8pt}
\caption{Fit for rho meson in terms of fragmentation functions with
(L) LEP data on the $Z$-pole and (R) HRS data at $\sqrt{s} = 29$ GeV.
The data\protect\cite{RhoZ1,RhoZ2,RhoZ3} at the $Z$-pole and from
HRS\protect\cite{RhoHRS} are shown with statistical and systematic errors
added in quadrature while the dots in the left-side plot show the fit
when averaged over the same $x$ bins as the data.}
\label{fig:rho}
\end{figure}

\paragraph{The $\rho, K^*$ cross-sections at the $Z$-pole}:
Given these fragmentation functions, finally we plot the various
cross-section combinations as given in Eq.~(\ref{eq:cross}) along with
data\cite{RhoZ1,RhoZ2,RhoZ3} for the $\rho$ meson in Fig.~\ref{fig:rho}.
The $\chi^2$ values of the fits (obtained by averaging the
cross-section over each bin and comparing with the data) are tabulated
in Table~\ref{tab:chi2}. (The difference between the average
cross-section and its value at the average $x$ value of the bin
indicated the uncertainty due to bin width and has been
included as an error in the computation. This affects the $\chi^2$
significantly only in the last bin).

We also use the fitted values to predict and compare with data on $\rho$
meson fragmentation at the photon-exchange-dominated regime of $\sqrt{s}
= 29$ GeV; see Fig.~\ref{fig:rho}. Note that the $\gamma$ exchange
process at lower energies is sensitive to a different combination of
the fragmentation functions than at the $Z$-pole.

\begin{figure}[pb]
%\centerline{\psfig{file=ijmpaf1.eps,width=4.7cm}}
\centerline{\includegraphics[angle=-90,width=0.49\textwidth]{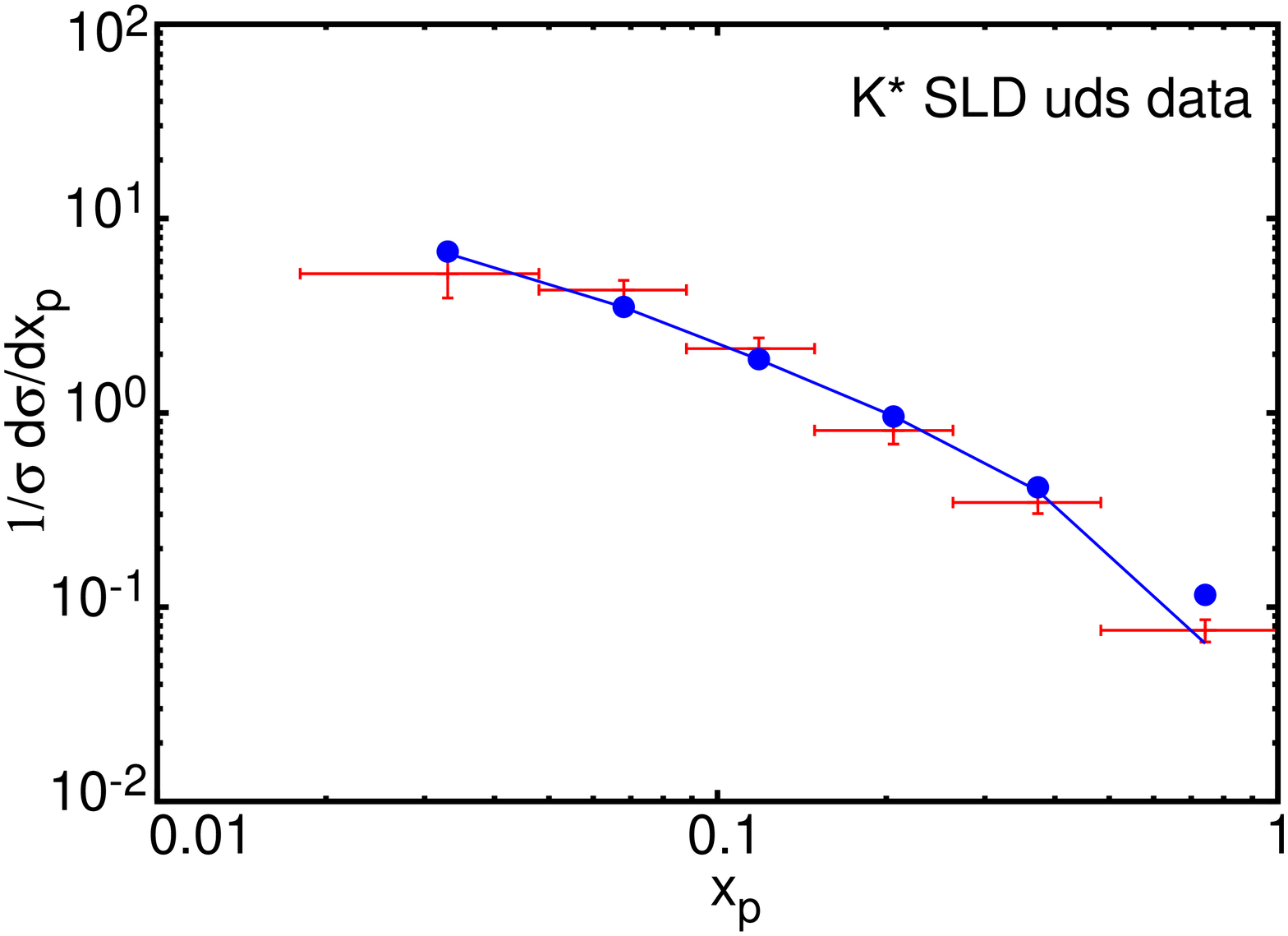}}
\vspace*{8pt}
\caption{Fit for $K^*$ meson with the best fit value of suppression
factor $\lambda = 0.063$.  Data\protect\cite{SLD1,SLD2} at the
$Z$-pole, from light quarks only, are also shown. See caption of
Fig.~\protect\ref{fig:rho} for more details.}
\label{fig:kstar}
\end{figure}

The value of the suppression factor can be determined by the fraction of
$\rho$ and $K^*$ meson at small $x$ (the data at large $x$ have relatively
larger error bars). A fit to the data as seen in Fig.~\ref{fig:kstar}
gives  $\lambda = 0.063$ (see Table~\ref {tab:unknown_par}).
Notice that the 1-$\sigma$ range of $\lambda$ ($0.05 \le \lambda \le 0.07)$ 
for the vector meson octet is close to that obtained $(\lambda \sim 0.08)$ 
for strangeness suppression of the pseudo scalar mesons $K^\pm, K^0$ and
$\overline{K^0}$. This may indicate that the origin of strangeness 
suppression may be independent of the spin structure of the mesons, 
since the two octets are otherwise unrelated.

The data\cite{SLD1,SLD2} for $K^*$ meson is also plotted along with our
predictions, using the same input parameters as used for the $\rho$
meson, in Fig.~\ref{fig:kstar} with the best fit value of the suppression 
factor for gluon, $f_g^{K^*} = 1.0$ (error bar is in
Table~\ref{tab:unknown_par}). The data are reasonably well fitted over
a large $x$ region. Thus all the parameters in the input fragmentation
functions are completely determined from fits to the $\rho$ and $K^*$
meson data. The $\rho$ data gives a set of $V$, $\gamma$ (and $D_g$),
which consistently fit the $K^*$ data with the inclusion of the two
parameters $\lambda$ and $f_g^{K^*}$.  Note that $\lambda$
contributes differently at low and high $x$: the suppression factors are
$(1+\lambda)/2$ and $\lambda$ for the valence and sea quark fragmentation
functions compared to those for $\rho$ fragmentation, as seen by a
comparison of Eqs.~(\ref{eq:D0rho}) and (\ref{eq:D0Kstar}). Hence the
excellent fits to the $\rho$ and $K^*$ meson data validate our simple
model in a non-trivial way.

Thus using the available data for $\rho$ and $K^*$ mesons we have fitted
the input fragmentation functions as listed in Table~\ref{tab:inputs}
using a simple broken SU(3) model. We now go on to apply the model in
the singlet-octet mixed $\omega$--$\phi$ sector, with the fragmentation 
functions $V, \gamma,$ and $\lambda$ fixed at the values obtained from 
fitting the $\rho, K^*$ data. 

\subsubsection{For mesons of octet and singlet mixture}
\label{analysis-mix}

As our model explains pure octet mesons in a nice fashion, we extended
it to mesons having mixture of octet and singlet components with same
fragmentation functions $V$, $\gamma$, $D_g$ and breaking parameter
$\lambda$. An additional fragmentation function expresses the singlet
contribution. However, this is simply related to a known octet
fragmentation function, as discussed earlier, and hence only a few
additional constant parameters are introduced in this extended sector,
namely $f_1^u, f_1^s, f_{sea}$ and $f_g$ for $\omega$ and $\phi$.

These describe the suppression in the sea as well as relate the
singlet fragmentation function to the octet one. With the help of
Eqs.~(\ref{du8}), (\ref{du1}) and (\ref{duop}) the fragmentation functions
for $\omega$ and $\phi$ mesons are expressed in terms of those of the
$\omega_8$ and $\omega_1$ mesons, including these unknown parameters. The
first is that of the mixing angle, $\theta$. It is known that
$\theta$ is large and positive\cite{Yao}, close to $35^\circ$. Exactly
at $35^\circ$, the $\omega$ meson is almost purely non-strange while
the $\phi$ is almost purely a strange $s\overline{s}$ hadron. 
A simultaneous best fit to $\omega$ and $\phi$ data (keeping the $V, 
\gamma$ and $\lambda$ fixed to the best fit values from the $\rho$ 
and $K^*$ analysis) gives $\theta = 42.6^\circ$ (error bar is in 
Table~\ref{tab:unknown_par}), not far from maximal. Hence the strange 
(non-strange) quark contribution to $\omega (\phi)$ is highly suppressed 
(the coefficients $(c_s^\omega)^2$  and $(c_u^\phi)^2$ are just a few
percent of $(c_u^\omega)^2$ and $(c_s^\phi)^2$ respectively.) Note the
data are inconsistent with no mixing, $\theta = 0^\circ$.

Hence the $\omega$ meson is totally dominated by $u$ and $d$ light quarks.
So it only has contribution from these two flavours whereas strangeness
has least contribution. Therefore, we expect this to be similar to $\rho$,
as is borne out by the similarity in cross-section behaviour. Since the
strange component of $\omega$ is highly suppressed, we fix the strange
singlet suppression factor to be $f_1^s = 0$ for $\omega$. We find the
data fit best to a sea suppression factor $f_{\rm sea}^\omega = 0.94$ and
$f_g^{\omega} = 1.0$ with $f_1^u$ consistent with zero (error bars are in
Table~\ref{tab:unknown_par}). This clearly shows that $\omega$ behaves
like $\rho$, with unsuppressed sea and gluon fragmentation functions
and very little contamination to the non-strange fragmentation functions
from the singlet mixing, that is, the singlet contribution merely serves
to make $\omega$ a practically non-strange meson orthogonal to $\rho$.

\begin{figure}[pb]
%\centerline{\psfig{file=ijmpaf1.eps,width=4.7cm}}
\includegraphics[angle=-90, width=0.49\textwidth]{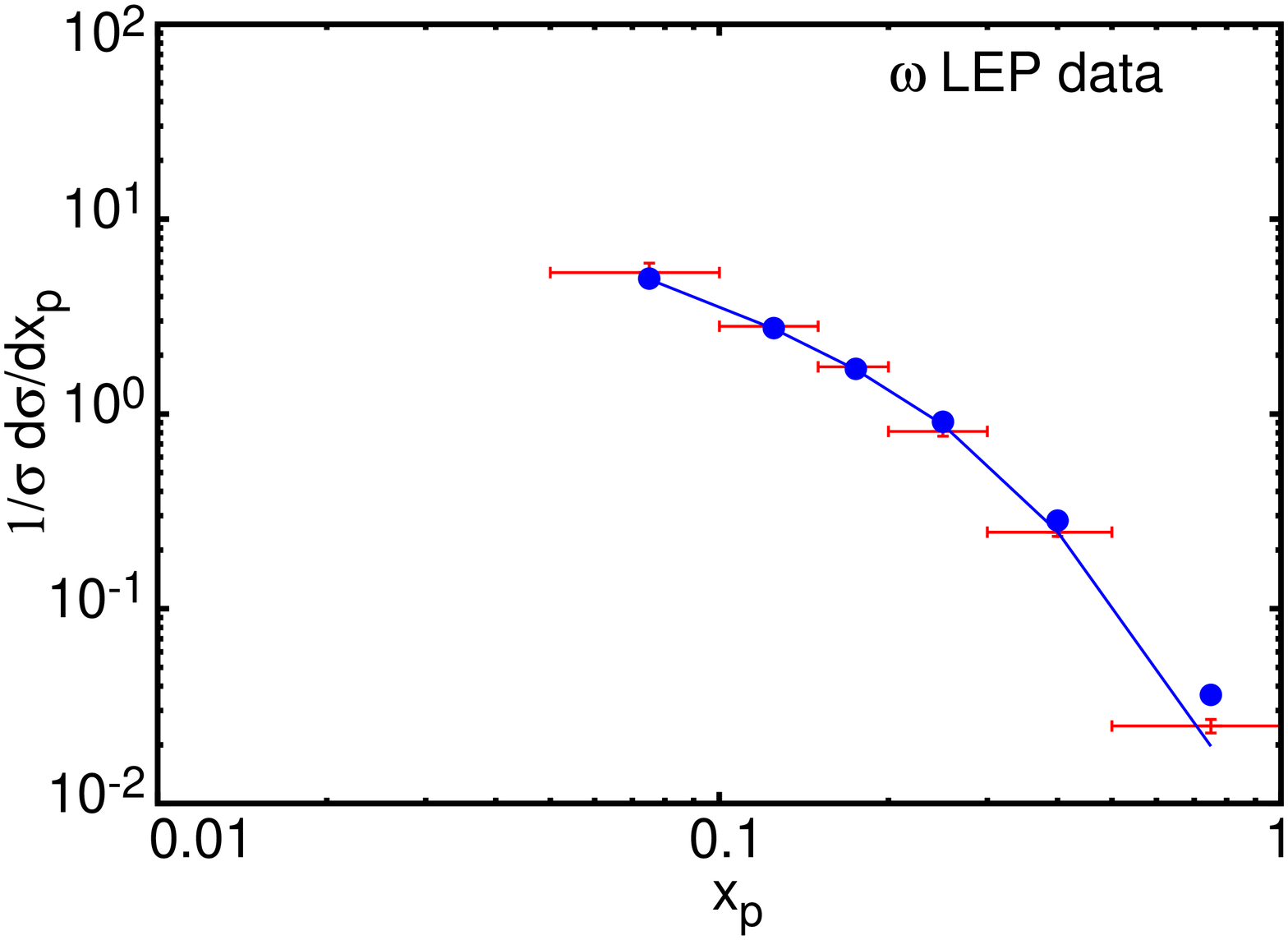}
\includegraphics[angle=-90, width=0.49\textwidth]{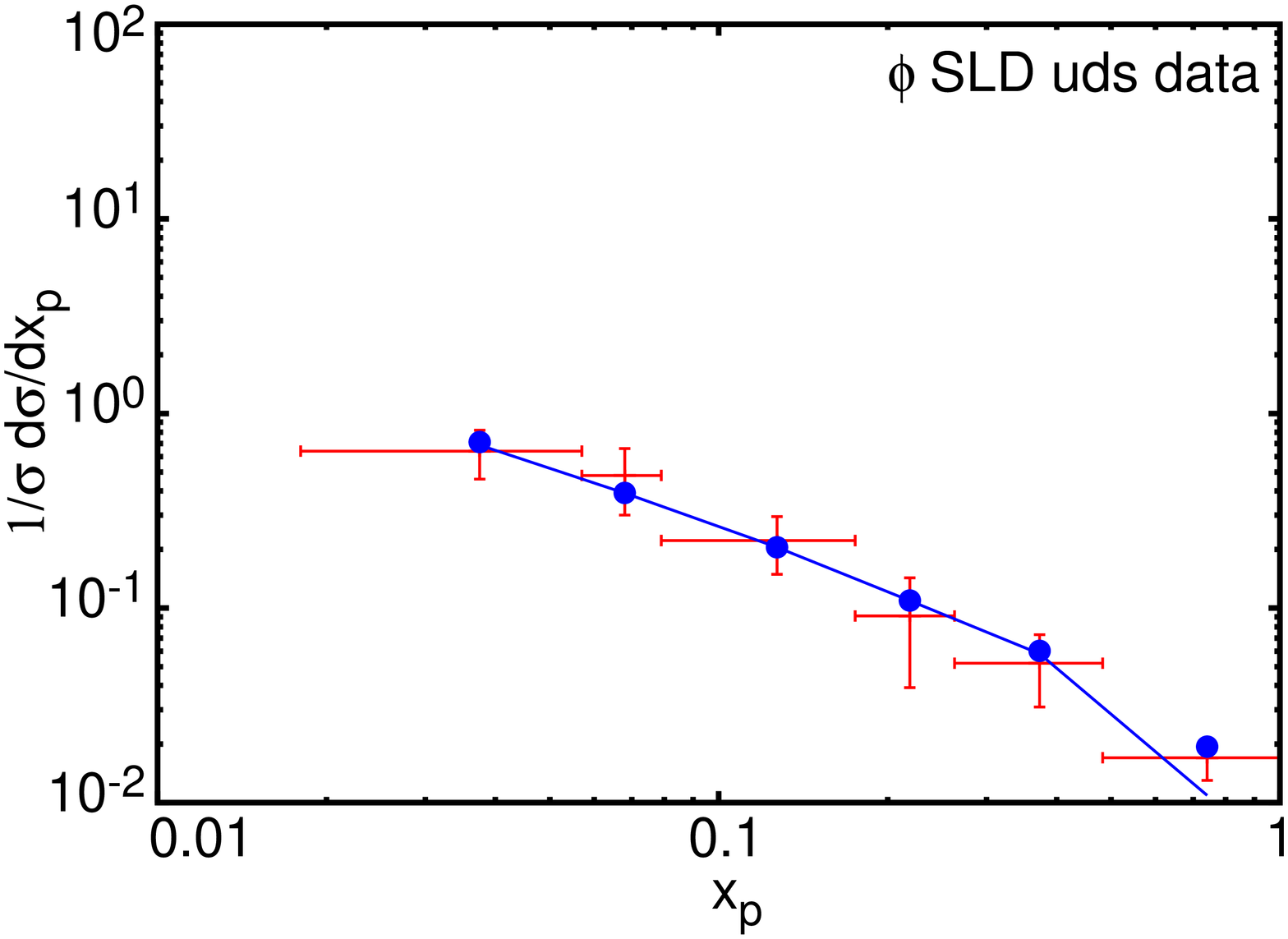}
\vspace*{8pt}
\caption{Fits to omega (L) and (R) phi meson. The data correspond to LEP
data\protect\cite{Omega1,Omega2} for $\omega$ and the SLD
data\protect\cite{SLD1,SLD2}
from light quarks alone for $\phi$. See the caption of
Fig.~\protect\ref{fig:rho} for more details.}
\label{fig:omegaphi}
\end{figure}

Similarly, we set $f_1^u = 0$ for $\phi$. Since $\phi$ is mostly 
saturated by the strange contribution, the sea fragmentation function
involves picking up both $s$ and $\overline{s}$ quarks and we therefore
expect the sea suppression factor to be of the order of $\lambda^2$.
Since the best-fit results were close to this value, with large error
bars, we simply set $f_{sea}^{\phi} = \lambda^2$. Meanwhile, the
best fit value of the singlet constant is $f_1^s = 4.0$.  The gluon
suppression factor for $\phi$ is $f_g^{\phi} = 0.32$ (error bars are
in Table~\ref{tab:unknown_par}). Hence, in contrast to the other member
of the nonet, $\phi$ is heavily suppressed both with respect to sea
quark and gluon fragmentation. These best-fit values result in a
cross-section behaviour as shown in Fig.~\ref{fig:omegaphi}.

A detailed analysis of the gluon
contribution and its relative suppression in $\phi$ mesons is best done
in a next-to-leading order (NLO) analysis where the gluon fragmentation
function directly appears in the expressions for the cross-section. An
alternative is to study hadro-production in
$p\,p$ scattering, where gluon fragmentation is equally dominant as
quark fragmentation due to the type of parton-level processes involved,
as stated earlier, although there are extra uncertainties due to
convolutions with parton density distributions in the cross-section
formulae. We therefore apply the fits obtained from hadro-production in
$e^+\,e^-$ processes, with a clean (non-hadronic) initial state, to
hadro-production in $p\,p$ collisions.

\subsection{Fragmentation in $p\,p$ process}
The PHENIX experiment at RHIC has measured $\omega$ and $\phi$
vector meson production\cite{RHICphi,RHICphipt1,RHICphipt2} in $p\,p$
collisions at $\sqrt{s} = 200$ GeV as a function of the transverse
momentum, $p_T$. We compute the relevant hadro-production cross-section
as expressed in Eqs.~(\ref{eq:pp}) and (\ref{eq:pTy}). We integrate over
a range $\pi$ in the azimuthal angle $\phi$ and over a rapidity range
$-0.35 \le y \le 0.35$ as stated earlier, and compare the cross-sections,
computed at a scale $Q^2 = p_T^2$, with the $p_T$-dependent data. We use
(GRV98-LO)\cite{GRV}, a standard set of parton distributions as available
in the CERN-libraries; a different choice of parton distributions will
not affect the results since the $(x,Q^2)$ range of the data are in the
well-studied range.  The data are binned in $p_T$ and the cross-sections
are quoted at the central value of the bin. As with $e^+\,e^-$ data,
the difference between the average cross-section and the cross-section
value at the average $p_T$ of the bin is included in the error while
computing the $\chi^2$ of the fits. Note that the fragmentation functions
are taken from the fits to the $e^+\,e^-$ data and there are no more
free parameters.

\begin{figure}[pb]
%\centerline{\psfig{file=ijmpaf1.eps,width=4.7cm}}
\includegraphics[angle=-90,width=0.49\textwidth]{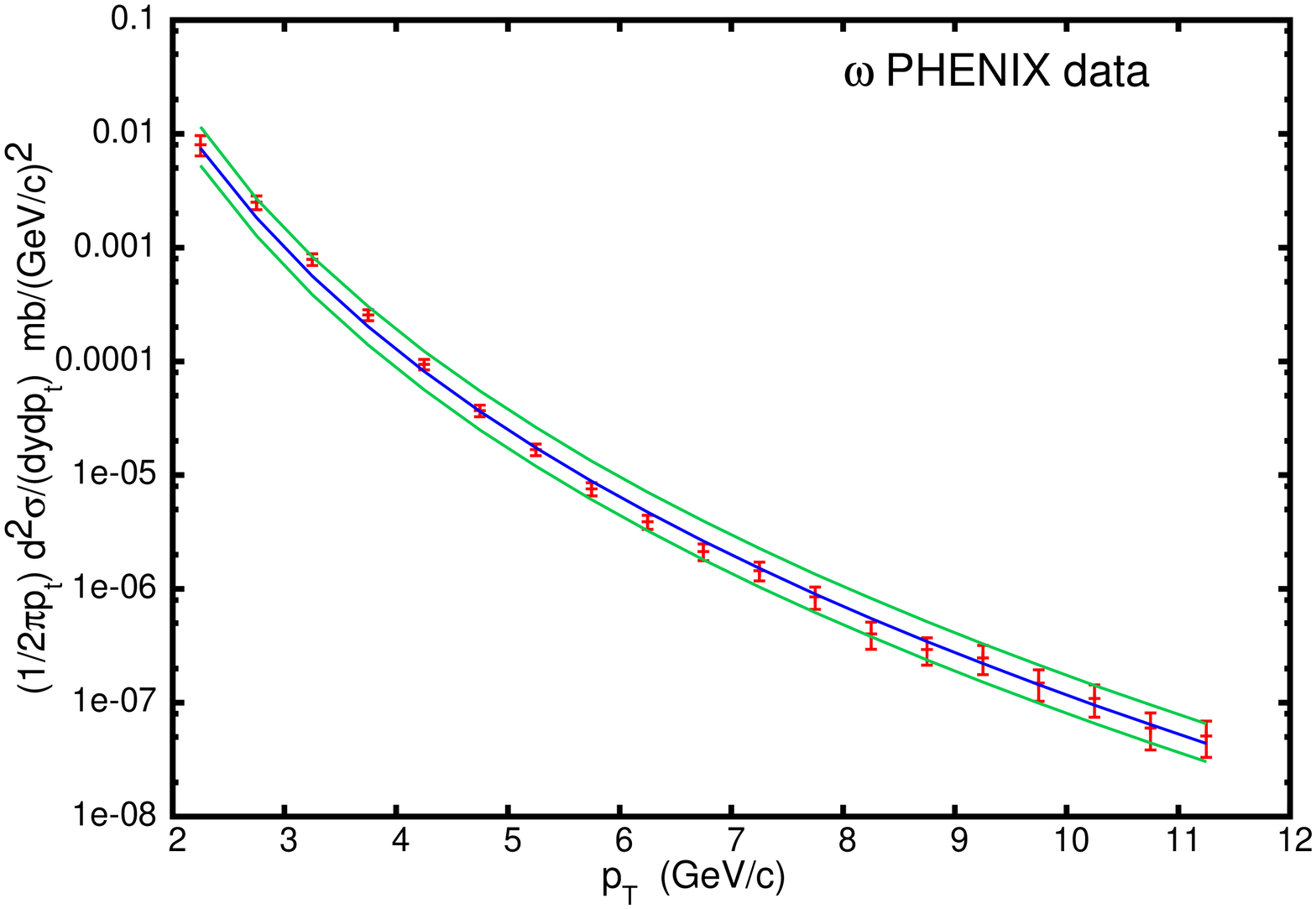}
\includegraphics[angle=-90,width=0.49\textwidth]{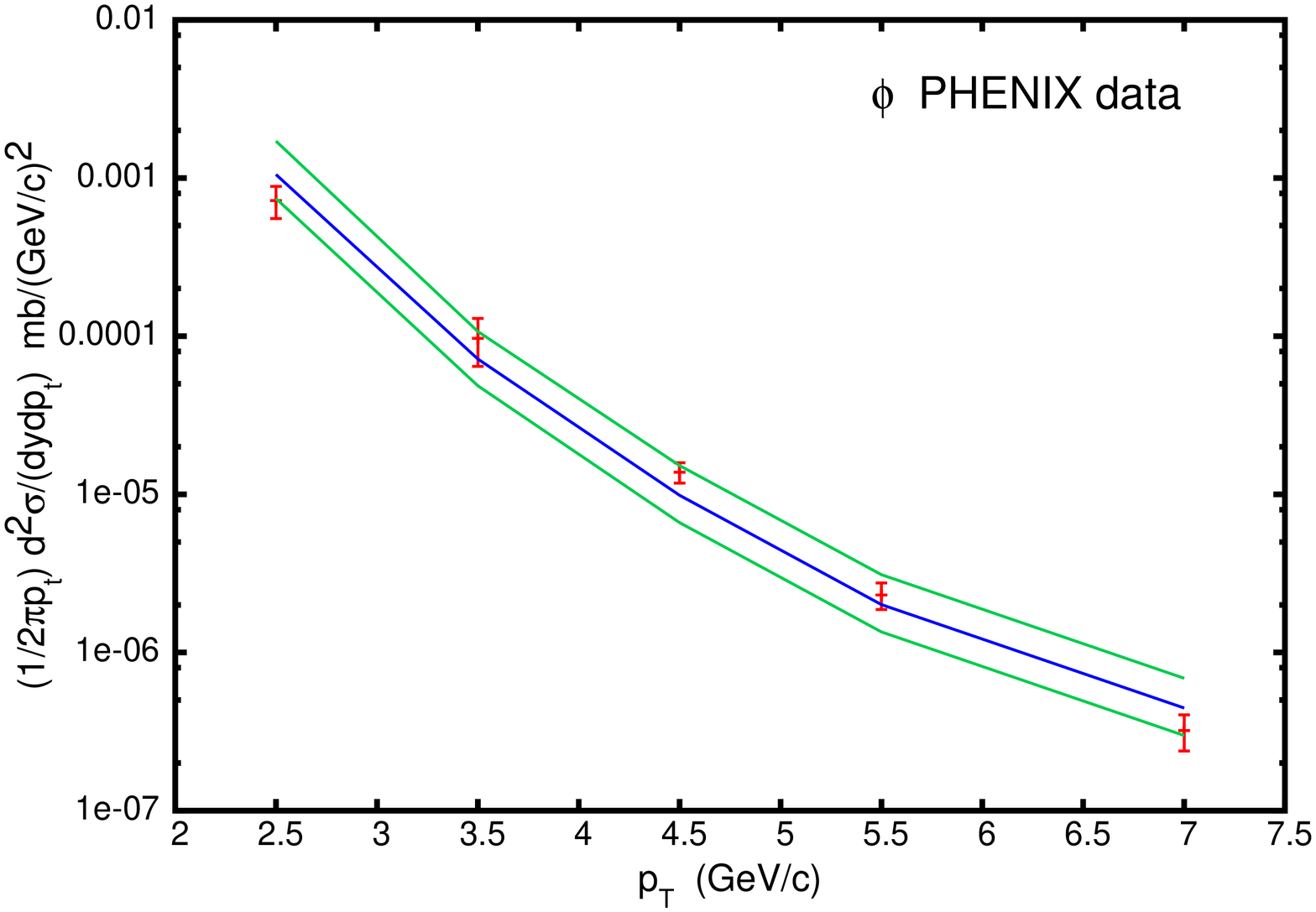}
\vspace*{8pt}
\caption{Cross section as a function of $p_T$ for omega (L) and phi
(R) meson hadro-production in $p\,p$ collisions. The data correspond to
PHENIX/RHIC data\protect\cite{RHICphi,RHICphipt1,RHICphipt2} for $\omega$
and $\phi$. Bands show the scale uncertainty on changing $Q^2=p_T^2$
over a range $p_T^2/2$ (upper curve) $\le Q^2 \le 2 p_T^2$ (lower curve).}
\label{fig:ppomegaphi}
\end{figure}

The results of the computation are shown for both $\omega$ and $\phi$
mesons in Fig.~\ref{fig:ppomegaphi} in comparison with data for which
the scale is larger than the starting scale of evolution,
$p_T^2>Q_0^2$. The figure also shows the band due
to the scale uncertainty over the range $p_T^2/2 \le Q^2 \le 2 p_T^2$.
It is seen that the model provides a good fit to the data. The values of
$\chi^2$ are listed in Table~\ref{tab:chi2} corresponding to the central
$Q^2$ value, $Q^2=p_T^2$.

\section{Summary and Discussion}
\label{summary}
Fragmentation functions of quarks and gluon for octet vector mesons
are studied for the first time in inclusive hadro-production in
electron-positron annihilation process and proton-proton collisions to
the leading order in QCD.

The unknown fragmentation functions were fitted to data on the $Z$
pole in $e^+\,e^-$ collisions. The model fairly depicts the octet
mesons $\rho$, $K^*$ and $\omega$ and greatly reduces the number of
unknown fragmentation functions since the symmetries reduce the 48 quark
fragmentation functions of the meson octet to combinations of just two:
the valence $V(x, Q^2)$ and sea $\gamma(x, Q^2)$ fragmentation functions,
apart from the gluon $D_g(x, Q^2)$ fragmentation function. This gives
the model great predictability, since the various (sparse) data sets
can be effectively combined to improve the quality of the fits.

An SU(3) breaking parameter $\lambda$ was introduced for the strange
$K^*$ mesons phenomenologically, to account for strangeness suppression,
at the starting scale of evolution.  The parameters at the input
scale were determined by comparison with data on the $Z$ pole. Heavy
quark contributions that are radiatively generated, are small and seen to
contribute at smaller $x$ values. The input distributions were tuned to
give good fits with inclusive hadro-production data at LEP for $\rho$
mesons and the ``pure uds" (tagged jet) $K^*$-production data at SLD
on the $Z$-pole. The best-fit values, along with the 1-$\sigma$ errors
are listed in Table~\ref{tab:inputs}. The best fit value of the
strangeness suppression factor is $0.05 \le \lambda \le 0.07$, close to
the similar suppression factor obtained in fits to the pseudoscalar octet
mesons, $\lambda \sim 0.08$, indicating that strangeness suppression in
quark fragmentation functions may be a spin-independent phenomenon.

The model was extended with some further assumptions to include a study
of the singlet vector meson with singlet--octet mixing (that is, to the
$\omega$ and $\phi$ mesons). While the mixing angle is known to be close
to $\theta \sim 35^\circ$, its value was determined through a simultaneous
best fit to the $\omega$ and $\phi$ data to be $\theta \sim 43^\circ$
(see Table~\ref{tab:unknown_par}). This is ultimately the focus of this
work since $\eta$ and $\phi$ hadro-production in $p\,p$ collisions are
used as a baseline when studying the production of these mesons as a
signal of quark gluon plasma in nucleus-nucleus collisions.

In order to describe the singlet sector, an ansatz was made, relating
the singlet fragmentation functions to one of the octet fragmentation
functions ($\alpha(x, Q^2)$). The constants of proportionality, $f_1^q$,
and the gluon suppression factors were determined through the fits.
While extending this model to include the octet and singlet nonet,
no new fragmentation functions were introduced: just a few more parameters
were included as described above, as well as the parameter $f_{sea}$ 
to describe suppression of sea quark fragmentation in both $\omega$ 
and (especially) $\phi$.

The proportionality constants $f_1^u$ and $f_1^s$ and the sea suppression
factor $f_{\rm sea}$ were determined by fitting the functions with the
data. Values of the parameters obtained appear to be phenomenologically
reasonable and are listed in Table~\ref{tab:unknown_par}, along with the
1-$\sigma$ errors, while the $\chi^2$ of the fits are given in
Table~\ref{tab:chi2}.

Note that the input fragmentation functions at a low energy scale, ($Q_0^2
= 1.5$ GeV$^2$) were included for three light flavours ($u$, $d$, $s$)
only. The heavier flavours charm and bottom are consistently included at
the appropriate thresholds during (leading order) evolution to the $Z$
pole. As is well known, next-to-leading order corrections will worsen
the fits at small-$x$ unless mass corrections are added. This is beyond
the scope of the present work.

Note also that the contribution at the $Z$-pole is dominated by the
flavour singlet fragmentation function $D_0$, which is the sum of the
individual quark flavour contributions with equal weight. Hence the fits
are most sensitive to this sum rather than to the individual fragmentation
functions. However, within the model, there are only two independent
fragmentation functions, a valence and a sea combination, in terms
of which {\em all} quark fragmentation functions are expressed. These
appear with different weights in the cross-sections for $\rho$ and $K^*$
production and hence can be individually determined.

At LO, the fits are rather insensitive to gluon fragmentation, although it
appears that the gluon fragmentation of $\phi$ is significantly suppressed
related to the others. Also, we have not included any isospin breaking
effects. In particular, charge asymmetries in fragmentation functions are
best studied through fragmentation in $e\,p$ scattering, while $p\,p$
processes are sensitive to the gluon fragmentation. However, data, as
well as analysis, is not as clean in this sector due to uncertainties in
scale, for instance, in the $p\,p$ case. Hence fits to the fragmentation
functions from data in $e^+\,e^-$ sector can then be used as constraints
while analyzing data from these other processes. Such an analysis was
done for $\omega$ and $\phi$ hadro-production in $p\,p$ collisions
and was found to be in good agreement with RHIC/PHENIX data. Note that
there are no free parameters in this fit since all the fragmentation
functions are determined from earlier fits to the LEP data. Reasonable
values of $\chi^2$ were obtained, as listed in Table~\ref{tab:chi2},
although the scale dependences are quite severe. This is expected to
improve at next-to-leading order, although again this is beyond the
scope of the present work.

In summary, fragmentation of the entire nonet of vector meson is
explained via a simple model with broken SU(3). The model includes
drastically few fragmentation functions and some constant parameters
at a low input scale which were then evolved to the scale of the data
(mainly at the $Z$-pole). While the fits to the pure octet vector mesons
were very good, reasonable fits were obtained when the model was extended
to study the mixed $\omega$ and $\phi$ mesons. The model continued to
give good fits to the hadro-production data in $p\,p$ collisions as well,
where the gluon fragmentation becomes important. This reflects the great
predictability and efficiency of this model, especially in view of the
paucity of data in this sector.

\section*{Acknowledgements}
We thank M V N Murthy for discussions and feed-back. One of the authors
HS would like to thank Professor A.S. Vytheeswaran, University of Madras,
for his motivation and suggestions. The author is also thankful to the
University of Madras for financial support in the form of a University
Research Fellowship (URF).

\begin{table}[ph]
%\tbl{Quark fragmentation functions into members of meson octet in terms of the SU(3) functions, $\alpha$, $\beta$ and $\gamma$.}
\caption{Quark fragmentation functions into members of meson octet in terms of the SU(3) functions, $\alpha$, $\beta$ and $\gamma$.}
{\begin{tabular}{ccl|ccl}
\toprule
fragmenting & \multicolumn{2}{c|}{${}_{\displaystyle K^{*+}}$} & fragmenting &
\multicolumn{2}{c}{${}_{\displaystyle K^{*0}}$} \\
quark & & & quark & & \\  \colrule
$u$ & : &  ${\alpha}+{\beta}+{\frac{3}{4}}{\gamma}$ & 
$u$ & : &  $2{\beta}+{\gamma}$ \\
$d$ & : &  $2{\beta}+{\gamma}$ & 
$d$ & : &  ${\alpha}+{\beta}+{\frac{3}{4}}{\gamma}$ \\
$s$ & : &  $2 {\gamma}$ & 
$s$ & : &  $2 {\gamma}$ \\ \colrule
fragmenting & \multicolumn{2}{c|}{${}_{\displaystyle \omega/\phi}$} & 
fragmenting &
\multicolumn{2}{c}{${}_{\displaystyle \rho^0}$} \\
quark & & & quark & & \\  \colrule
$u$ & : &  
$\frac{1}{6}{\alpha}+\frac{9}{6}{\beta}+\frac{9}{8}{\gamma}$ &
$u$ & : &  
$\frac{1}{2}{\alpha}+\frac{1}{2}{\beta}+\frac{11}{8}{\gamma}$ \\
$d$ & : &  
$\frac{1}{6}{\alpha}+\frac{9}{6}{\beta}+\frac{9}{8}{\gamma}$ &
$d$ & : &  
$\frac{1}{2}{\alpha}+\frac{1}{2}{\beta}+\frac{11}{8}{\gamma}$ \\
$s$ & : &  
$\frac{4}{6}{\alpha}+\frac{9}{6}{\gamma}$ &
$s$ & : &  $2\beta+\gamma$ \\ \colrule
fragmenting & \multicolumn{2}{c|}{${{}_{\displaystyle \rho^+}}$} &
fragmenting &  \multicolumn{2}{c}{${{}_{\displaystyle \rho^-}}$} \\
quark & & & quark & & \\  \colrule
$u$ &  : & ${\alpha}+{\beta}+{\frac{3}{4}}{\gamma}$ & 
$u$ &  : & $2 {\gamma}$ \\ 
$d$ &  : & $2 {\gamma}$ & 
$d$ &  : & ${\alpha}+{\beta}+{\frac{3}{4}}{\gamma}$ \\ 
$s$ &  : & $2{\beta}+{\gamma}$ & 
$s$ &  : & $2{\beta}+{\gamma}$ \\ \colrule
fragmenting & \multicolumn{2}{c|}{${{}_{\displaystyle \overline{K^{*0}}}}$} &
fragmenting & \multicolumn{2}{c}{${{}_{\displaystyle K^{*-}}}$} \\
quark & & & quark & & \\  \colrule
$u$ & : & $2{\beta}+{\gamma}$ & 
$u$ & : & $2 {\gamma}$ \\ 
$d$ & : & $2 {\gamma}$ & 
$d$ & : & $2{\beta}+{\gamma}$ \\ 
$s$ & : & ${\alpha}+{\beta}+{\frac{3}{4}}{\gamma}$ & 
$s$ & : & ${\alpha}+{\beta}+{\frac{3}{4}}{\gamma}$ \\ \botrule
\end{tabular}
\label{tab:frag}}
\end{table}

\vskip 2truecm
\begin{table}[ph]
%\tbl{Best fit values of the parameters defining the input fragmentation functions at the starting scale of $Q^2 = 1.5 \ GeV^2$ , with their 1-$\sigma$ error bars.}
\caption{Best fit values of the parameters defining the input fragmentation functions at the starting scale of $Q^2 = 1.5 \ GeV^2$ , with their 1-$\sigma$ error bars.}
{\begin{tabular}{ccccc} \toprule
 &      & Central Value   & \multicolumn{2}{c}{Error Bars} \\ \colrule

$V$     &$a$ & 0.66   & -0.07    & 0.07    \\  
        &$b$ & 0.52   & -0.12    & 0.32    \\  
        &$c$ & 1.48   & -0.13    & 0.15    \\  
        &$d$ & 4.54   & -0.50    & 0.50    \\  
        &$e$ & -3.29  & -0.91    & 0.92    \\  
$\gamma$&$a$ & 1.10   & -0.05    & 0.05    \\  
        &$b$ & -0.31  & -0.03    & 0.03    \\  
        &$c$ & 7.42   & -0.15    & 0.15    \\  
        &$d$ & 3.57   & -0.62    & 0.62    \\  
        &$e$ & 20.81  & -2.44    & 2.49    \\  
$D_g$   &$a$ & 1.91   & -0.25    & 0.25    \\  
        &$b$ & 2.82   & -0.19    & 0.22    \\  
        &$c$ & 3.40   & -0.14    & 0.16    \\  
        &$d$ & 8.54   & -1.36    & 1.36    \\  
        &$e$ & 0.00   &  --    & --    \\ \botrule
\end{tabular}
\label{tab:inputs}}
\end{table}

\vskip 2truecm
\begin{table}[ph]
%\tbl{Best fit values of the parameters defining the input fragmentation functions at the starting scale of $Q^2 = 1.5 \ GeV^2$, with their 1-$\sigma$ error bars.}
\caption{Best fit values of the parameters defining the input fragmentation functions at the starting scale of $Q^2 = 1.5 \ GeV^2$, with their 1-$\sigma$ error bars.}
{\begin{tabular}{ccccc} \toprule
 &      Central Value   & \multicolumn{2}{c}{Error Bars} \\ \colrule
$\lambda$          & 0.063   & -0.01  & 0.01   \\  
$\theta $          & 42.6  & -2.0  & 2.0   \\  
$f_{sea}^{\omega}$ & 0.94   & -0.08  & 0.08   \\  
$f_1^u(\omega)$    & 0.1    & -0.3  & 0.3  \\  
$f_1^s(\phi)$      & 4.0   & -1.4  & 1.5   \\  
$f_g^{K^*}$        & 1.0   & -0.2  & 0.2   \\  
$f_g^{\omega}$     & 1.0   & -0.7  & 0.7   \\  
$f_g^{\phi}$       & 0.32   & -0.08  & 0.09   \\  \botrule
\end{tabular}
\label{tab:unknown_par}}
\end{table}

\vskip 2truecm
\begin{table}[ph]
%\tbl{$\chi^2$ for fits to inclusive vector meson production data from $e^+\,e^-$ experiments on the $Z$-pole from LEP and SLD experiments and from $p\,p$ experiments from PHENIX at RHIC.}
\caption{$\chi^2$ for fits to inclusive vector meson production data from $e^+\,e^-$ experiments on the $Z$-pole from LEP and SLD experiments and from $p\,p$ experiments from PHENIX at RHIC.}
{\begin{tabular}{ccccc} \toprule
Data Set           & No. of data points  & $\chi^2$     \\   \colrule
$\rho$ (ALEPH)      &   8                &     4.8      \\  
$\rho$ (DELPHI '95) &   6                &     2.0      \\  
$K^{*0}$ (SLD)     &    6                &     7.1      \\  
$\omega$ (ALEPH)   &    6                &     0.7      \\  
$\phi$ (SLD)       &    6                &     1.0      \\ \botrule
$\omega$ (PHENIX)  &    19               &     21.4      \\  
$\phi$ (PHENIX)    &    5                &     5.3      \\ \botrule

\end{tabular}
\label{tab:chi2}}
\end{table}

\end{document}